\documentclass[a4paper, 10pt, conference]{ieeeconf}      

\IEEEoverridecommandlockouts                              
\usepackage{graphicx}
\usepackage{amsmath}
\usepackage{amssymb}
\usepackage{subcaption}
\usepackage{color,soul}
\usepackage{multirow}
\usepackage{hyperref}
\usepackage{graphics} 
\usepackage{epsfig} 
\usepackage{mathptmx} 
\usepackage{times} 
\usepackage{gensymb}

\title{\LARGE \bf End-to-end sensor modeling for LiDAR Point Cloud}

\author{Khaled Elmadawi$^{1}$, Moemen Abdelrazek$^{1}$, Mohamed Elsobky$^{1}$, Hesham M. Eraqi$^{1}$, and Mohamed Zahran$^{1}$
\thanks{$^{1}$Khaled Elmadawi, Moemen Abdelrazek, Mohamed Elsobky , Hesham M. Eraqi, and Mohamed Zahran are with Valeo AI Research, Cairo, Egypt {\tt\small khaled.elmadawi@gmail.com, \{moemen.abdelrazek.ext, mohamed.elsobky, hesham.eraqi, mohamed.zahran\}@valeo.com}}%
}

\begin{document}

\maketitle
\thispagestyle{empty}
\pagestyle{empty}

\begin{abstract}
Advanced sensors are a key to enable self-driving cars technology. Laser scanner sensors (LiDAR, Light Detection And Ranging) became a fundamental choice due to its long-range and robustness to low light driving conditions. The problem of designing a control software for self-driving cars is a complex task to explicitly formulate in rule-based systems, thus recent approaches rely on machine learning that can learn those rules from data. The major problem with such approaches is that the amount of training data required for generalizing a machine learning model is big, and on the other hand LiDAR data annotation is very costly compared to other car sensors. An accurate LiDAR sensor model can cope with such problem. Moreover, its value goes beyond this because existing LiDAR development, validation, and evaluation platforms and processes are very costly, and virtual testing and development environments are still immature in terms of physical properties representation.

In this work we propose a novel Deep Learning-based LiDAR sensor model. This method  models the sensor echos, using a Deep Neural Network to model echo pulse widths learned from real data using Polar Grid Maps (PGM). We benchmark our model performance against comprehensive real sensor data and very promising results are achieved that sets a baseline for future works.
\end{abstract}

\section{Introduction}
  In this paper we are introducing a novel method to mimic LiDAR sensor from reality into simulation environments by using Deep Neural Networks, giving us a chance to generate as much data as possible with a limited cost, unlike the followed methods used in the automotive field as will be explained in the below sections.
 However, before explaining our problem and our proposed solution, let's start establish some basics and definitions about LiDAR sensor, how it works and what exactly are we modeling. 
 \subsection{LiDAR Sensor}
\begin{figure}[thpb]
  \centering
  \framebox{\includegraphics[scale=0.28]{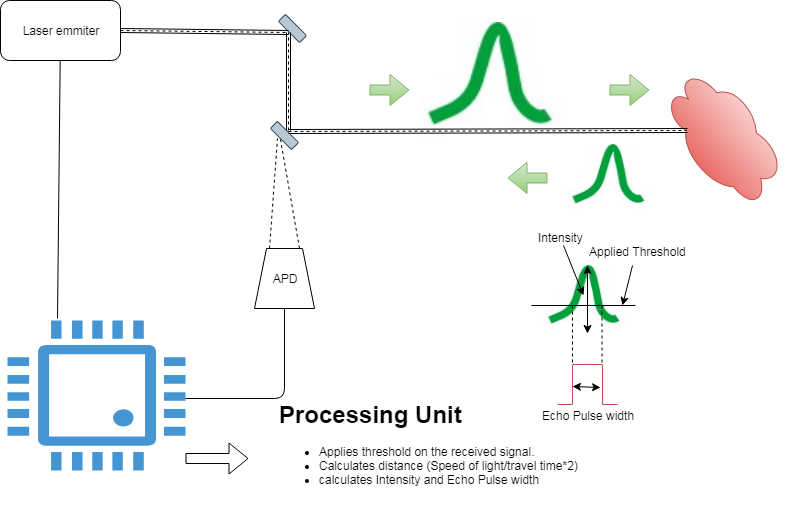}}
  \caption{\label{fig:LidarAnatomy} Basic Lidar Anatomy.}
\end{figure}

LiDAR sensor basic components are emitter, receptor(Avalanche Photodiode known as APD), and a processing unit. LiDAR sensor measures distance by multiplying the speed of light by half the time of travel of the pulse from the emitter to the receiver APD.LiDAR processing unit applies filters and thresholds on the received pulse to differentiate between noisy detections and meaningful detections. Received LiDAR pulse carries information about the materials and the reflectively of the detected object. These information are represented in one of two values, detected reflection Intensity, or detected reflection Echo Pulse width depending on what exactly the LiDAR processing unit is calculating or returning in it's byte stream, as shown in fig [\ref{fig:LidarAnatomy}].

\begin{figure}[thpb]
  \centering
    \framebox{\includegraphics[scale=0.48]{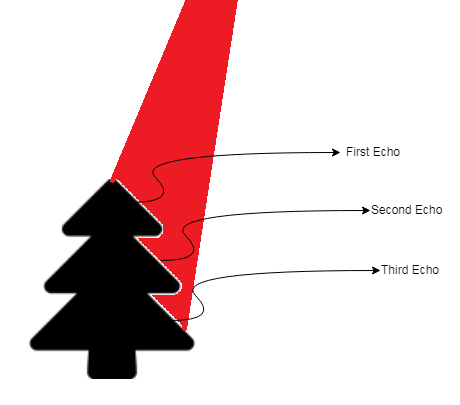}}
  \caption{\label{fig:LidarMultipleEchos} LiDAR multiple echos phenomina.}
\end{figure}

LiDAR sensor can have multiple reflections from the same pulse due to the division of the ray profile resulting into multiple echos as shown in fig [\ref{fig:LidarMultipleEchos}].

In our work we are mimicking the EPW from real LiDAR sensor, with it's multiple reflections as will be illustrated in the coming sections.


\subsection{Motivation}

Data generation\cite{geiger2013vision} in Self-driving cars applications is one of the biggest challenges that manufacturers are facing in the field, where it is either gathering data from the street, and annotating them, then passing them through development and validation processes and cycles which is too expensive, or generating syntactic data from an artificial environment that will be very ideal and far from realistic effects.

\subsection{Contribution}

\begin{figure}[thpb]
  \centering
    \framebox{\includegraphics[width=\columnwidth]{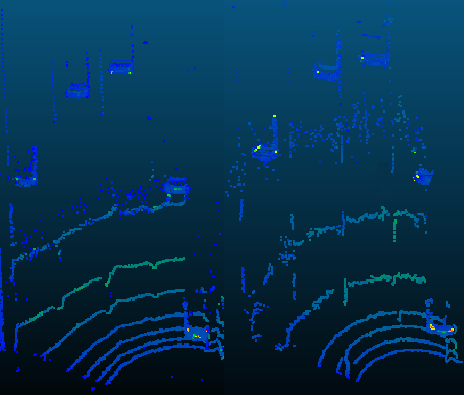}}
    \caption{\label{fig:realVsSim} A comparison between real LiDAR data and data from synthetic data generated from our proposed sensor model to left and right respectively. Each scan point color represent its Echo Pulse Width (EPW) value. It is obvious that both examples 
    1- Our approach has clearly mimicked EPW values from real data. 
    2- Our approach could mimic noise model in syntactic generated data in the far perception.
    3- Our model could learn how to represent lanes as learned from real traces}
\end{figure}

We provide a measurable accurate sensor model that represents LiDAR sensors physical properties in EPW and noise model. Our proposed sensor model can run in real-time and compatible with the famous simulation environments used in automotive industry (like CarMaker  [\ref{fig:Carmakerdemo}] , Unity [\ref{fig:Unitydemo}] and ROS[\ref{fig:Summary}]). We conduct a quantitative and qualitative benchmarks against comprehensive real LiDAR data which demonstrate the effectiveness of our proposed sensor model, and present measurable way of how close or far are we from the real sensor model itself, and to give you a closer image of what we are presenting, we challenge you to identify which of the two  images in fig [\ref{fig:realVsSim}] is coming from simulation environment, and which is coming from real environment.

\section{Related Work}
\label{sec:related}
In this section we provide the related work to real to sim and sim to real transformations, where we can use simulation environments  in to train a reinforcement learning DNN then applying it in real world, as shown in sim-to-Real\cite{matas2018sim}, or by generating data streams from syntactic environment to provide more training data to the neural networks, provide the neural network with scenarios that can't be reproduced in real life, and having an automatic annotation to the generated perception\cite{gaidon2016virtual}, all of these efforts require the syntactic generated environments to be mapped from the ideal simulated domain to real domain with contains imperfection and physical effects or from domain A to domain B as shown in \cite{isola2017image} \cite{shrivastava2017learning}, otherwise self driving vehicles will suffer a huge bias in the performance when we shift from simulation domain to real domain.

This is what we propose in this paper, we propose a method of how to transform LiDAR perception from simulated ideal domain to real domain by using end to end Deep neural network stack that contains physical properties, environmental effects and sensor artifacts as shown in fig [\ref{fig:realVsSim}] 

\section{Our Approach}
\label{sec:method}
\begin{figure}[thpb]
  \centering
      \framebox{\includegraphics[scale=0.235]{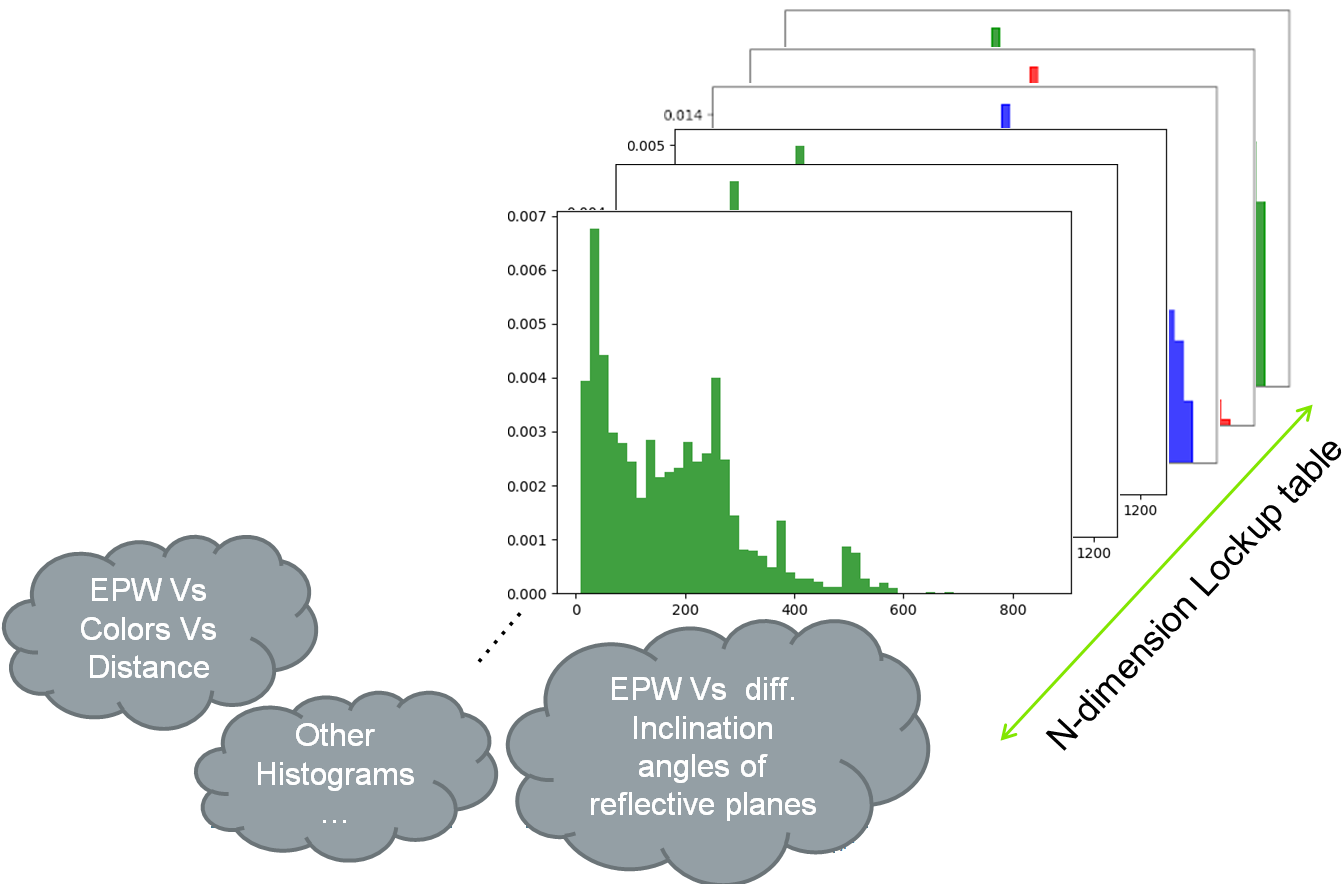}}
    \caption{\label{fig:MultidimensionalLockuptable} Multidimensional Lockup table that the DNN need to learn.}
\end{figure}

We hypothesize that the EPW value depends mainly on the object materials, distance from the sensors, and the laser beam inclination and yaw shooting angles. Hence we construct two kind of histograms; one for EPW values per echo and object materials, and the other one is the probability of echos occurrence over different yaw shooting angles. Such histograms as shown in fig [\ref{fig:MultidimensionalLockuptable}] represent a look-up table that contains EPW sensitivity to aforementioned dependencies, which relate it to signal strength or attenuation, noise model through different distances, different objects and echos occurrence distributions information. Typically, as will be further detailed in the Experimental Setup section, the sensor we used in our analysis supports up to 3 echos. Our hypothesis validity is verified in the experiments and results section. 

In addition, we propose an end-to-end deep learning-based approach that learns the complex function mapping from physical environmental setup to LiDAR data physical properties.

\begin{figure}[thpb]
  \centering
    \framebox{\includegraphics[scale=0.27]{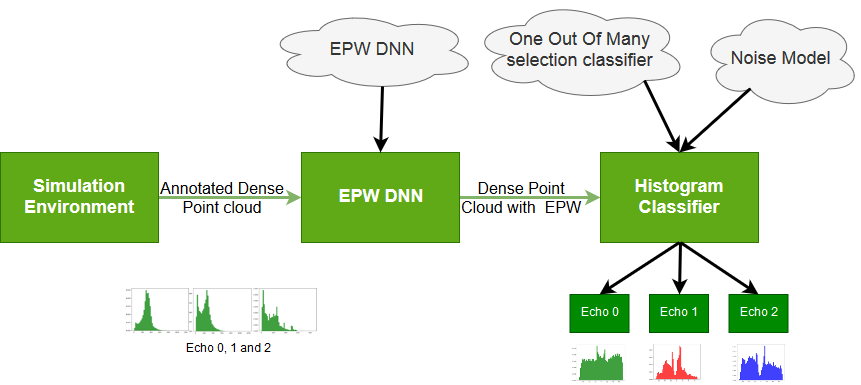}}
    \caption{\label{fig:Solution_pipeline} DNN Pipeline that Encapsulate Sensor model N-dimensional Lockup table.}
\end{figure}

Our architecture is two stage DNN, first is fully convolutional DNN that inferences EPW value based on Polar grid map approach, second is one out of many selection block, where it selects one discrete representation of the ray profile of the many discrete signals of the simulated ray profile for each echo, and from this selection noise model is represented, as shown in fig [\ref{fig:Solution_pipeline}].

\subsection{Architecture}


Our architecture is based mainly on two stages, first is fully convolutional DNN that inferences EPW value based on Polar grid map approach, second is one out of many selection block, where it selects one discrete representation of the ray profile of the many discrete signals of the simulated ray profile, and from this selection noise model is applied.

\subsection{Polar Grid Map}
half of the solution in most of the deeplearning problems is in the approach that the problem is tackled by, how will you represent the input, the objective that is required to be learned by the DNN, and what exactly does the output layer represents.  

\begin{figure}[thpb]
  \centering
  \framebox{\includegraphics[width=\columnwidth]{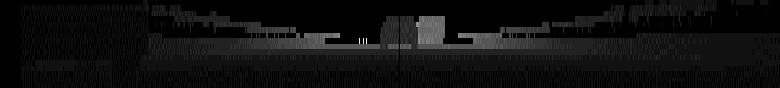}}
     \framebox{\includegraphics[width=\columnwidth]{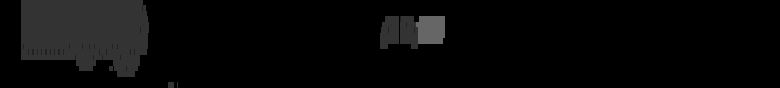}}
    \caption{\label{fig:PGM} Annotated Polar Grid Map point cloud, Upper PGM is depth representation, lower PGM is point level annotation.}
\end{figure}

The Polar Grid Map (PGM) is a representation for a LiDAR full scan in a 3D tensor. The full scan is composed of a cloud of scan points which is encoded in a PGM. Each channel of the PGM is a 2D grid map, where each row represents a horizontal LiDAR layer. Having the sensor as a reference point, each scan point is determined by a distance from such reference, and azimuth and altitude angles. Each PGM cell corresponds to a scan point, the row and column indices represent the scan point altitude and azimuth angles respectively. The cell value represent information about the corresponding scan point. In the first PGM channel, the value holds the scan point distance, while the second channel holds its class. The same representation can be extended to represent more information by adding extra channels, while in our study we only use the two mentioned channels.
An example for PGM representation coming from a single LiDAR full scan is shown in fig [\ref{fig:PGM}].

\begin{figure}[thpb]
  \centering
    \framebox{\includegraphics[scale=0.17]{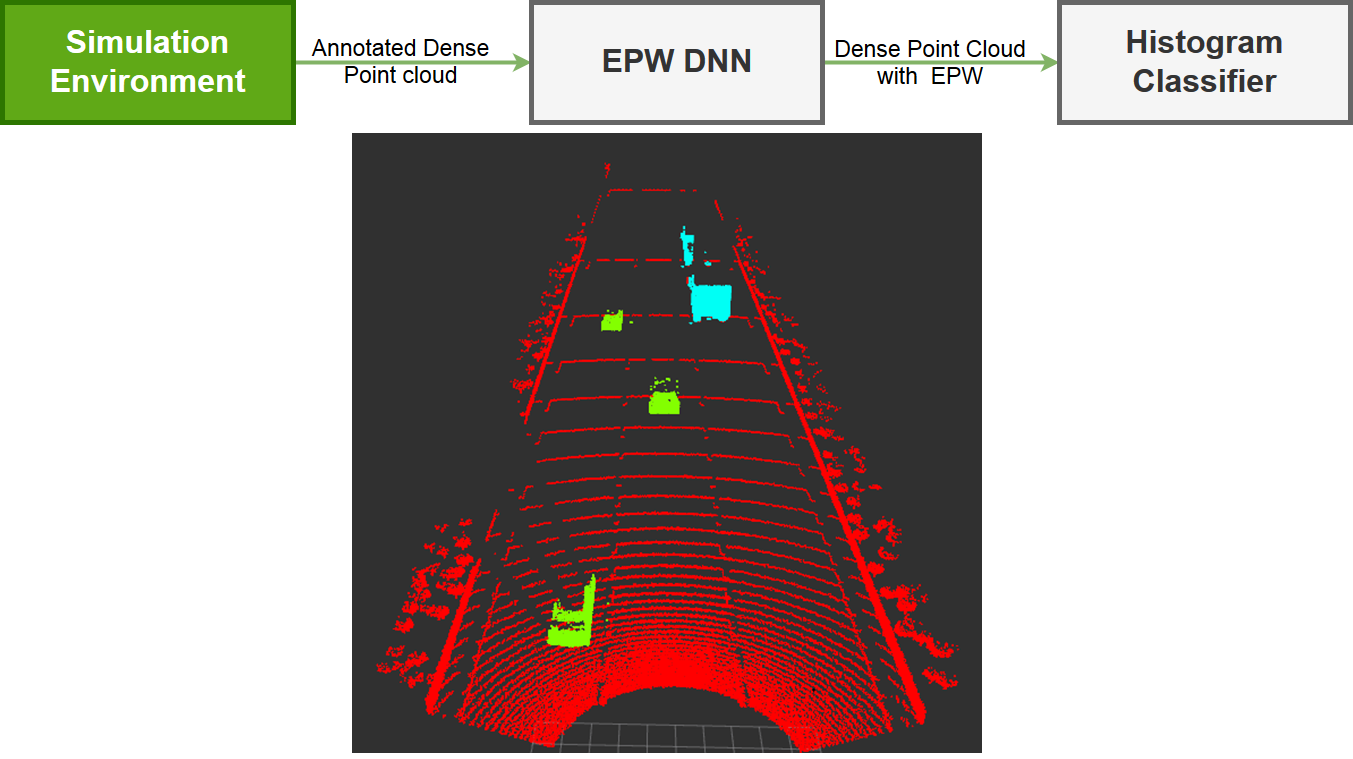}}
    \caption{\label{fig:InputFormat} Dense Point level Annotated Point Cloud.}
\end{figure}

The input coming from our simulation environments would be a dense annotated point cloud, the dense point cloud represents the discrete representation of the laser ray profile, as shown in fig [\ref{fig:InputFormat}].

\subsection{EPW DNN}

\begin{figure}[thpb]
\centering
\framebox{\includegraphics[scale=0.215]{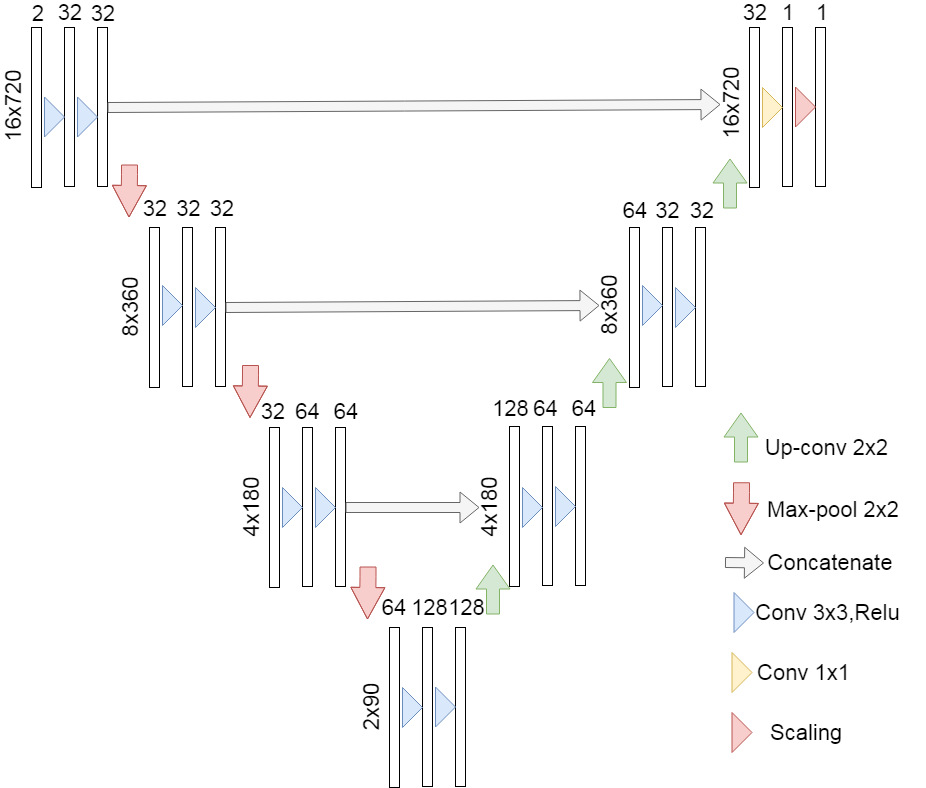}}
    \caption{\label{fig:unet} Unet architecture. Each white
box corresponds to a multi-channel feature map. The number of channels is denoted
on top of the box. The x-y-size is provided at the middle of the box.}
\end{figure}

We adopt the U-Net architecture \cite{ronneberger2015u} to work on input data of only two channels, and we reduce the number of blocks to fulfill reasonable run-time constraints. For each full scan, the corresponding PGM is used as an input to the network. It goes through an encoder composed of three down-sampling blocks, each block has two convolutional layers with variable number of 3x3 kernels, followed by a max-pooling layer. Then the output goes through three up-sampling blocks, each block has 2 convolutional layers with variable number of 3x3 kernels followed by a transpose convolution layer. The contracting paths, or skip connections as in \cite{ronneberger2015u}, from U-Net are adopted to capture LiDAR data context. The network output represents a PGM of 1 channel that holds the EPW information for the input full scan. The network is shown in fig[\ref{fig:unet}]. 

we use reduced mean square error in the epw dnn loss with regularization.We chosen this loss to penalize more large errors.
\begin{equation}
Loss = \frac{1}{n}\sum_{i=1}^{n}(y_i-y'_i)^2 + \frac{\lambda}{2}*\sum_{}^{}W_t
\end{equation}

\begin{figure}[thpb]
  \centering
  \framebox{\includegraphics[width=\columnwidth]{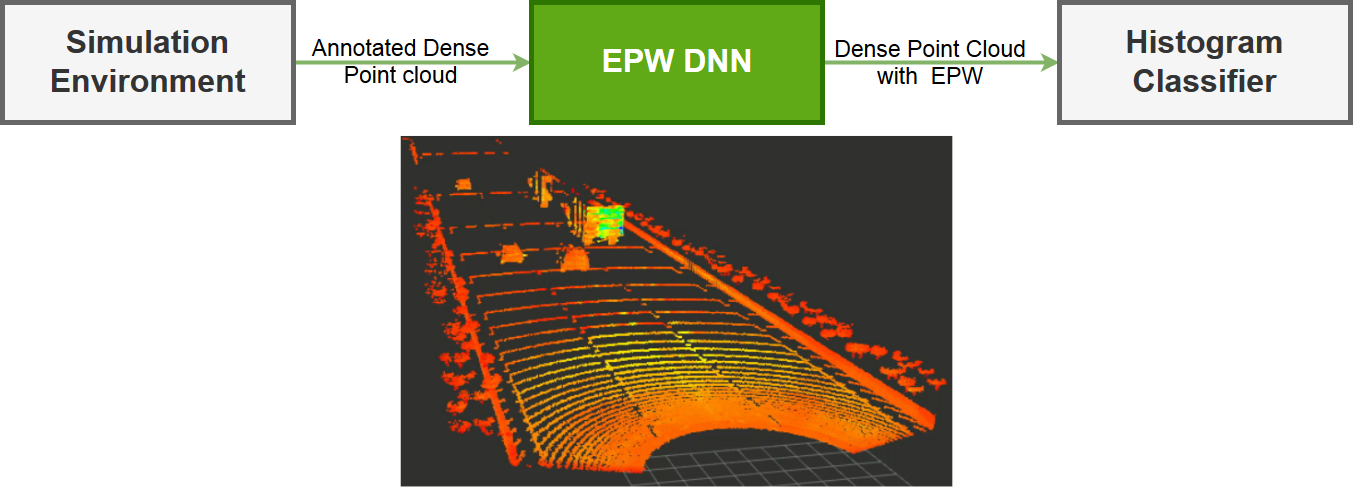}}

    \caption{\label{fig:1ststageOutput} Point cloud with its inferred EPWs.}
\end{figure}

The EPW DNN output's is the same point level annotated point cloud with their Inferred EPW, as shown in fig [\ref{fig:1ststageOutput}].

The EPW was trained in 350 epochs, with early stopping depending on the L1 score, we used a batch size of 8 samples, learning rate was  1e-5, and the annotated classes were [None, Cars, Trucks, Pedestrians, Motorbikes, High reflective materials].

\subsection{Second Stage Histogram Classifier}
\begin{figure}[thpb]
  \centering
    \framebox{\includegraphics[width=\columnwidth]{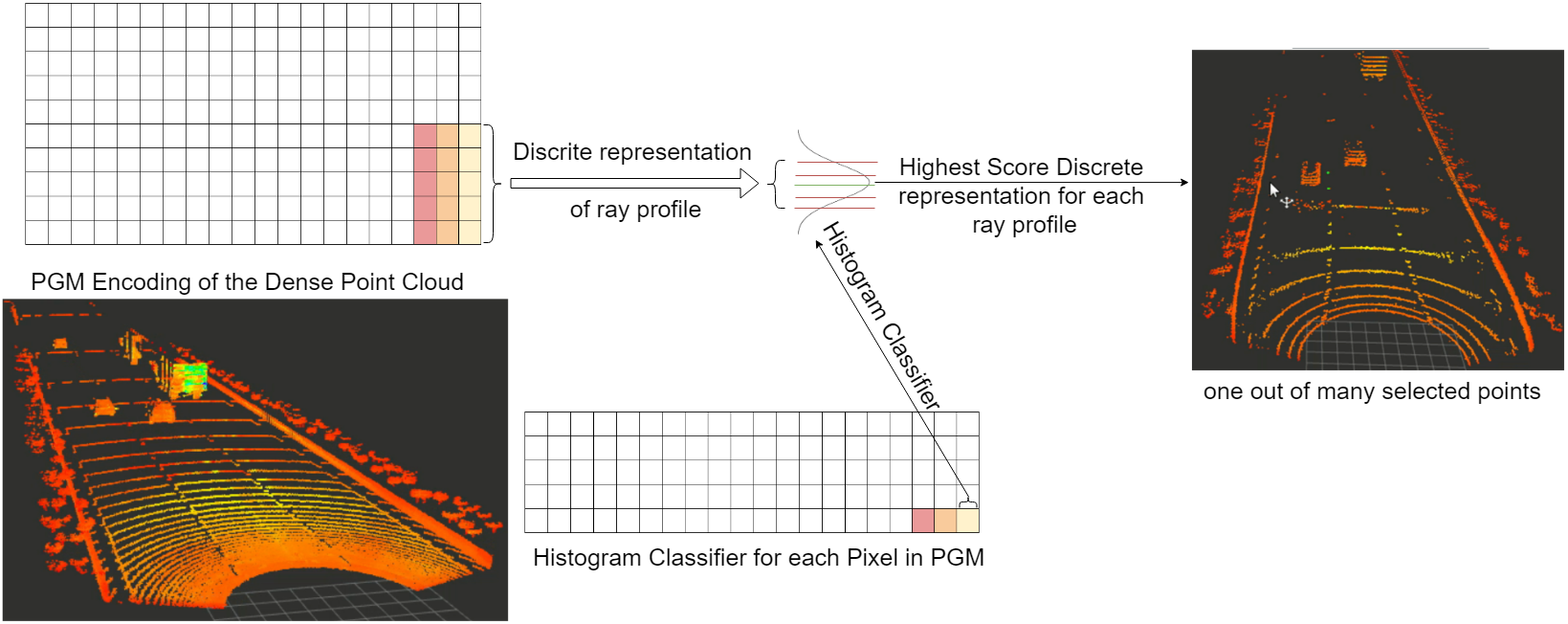}}
    \caption{\label{fig:HistogramClassifier} Histogram Bayes Classifier output, one out of many selection block.}
\end{figure}

We pass the Dense Point cloud to a Histogram Classifier to make one out of many point selection, where it selects from the Dense Point cloud what makes the Scan points looks like the realistic scan points as in real Lidar sensor model, as shown in fig [\ref{fig:HistogramClassifier}].

\section{Experiments and Results}
\label{sec:exps}

\subsection{Experimental Setup}
\subsubsection{Lidar Specs}:

Lidar is a 16 layer sensor $\pm 5$\degree    with vertical resolution of 0.625\degree, with 145\degree horizontal FOV with resolution that of 0.125\degree. 
The sensor provides three echos for each reflected ray cast, we may have for each reflection one, two or three scan point(s) based on the reflectivity of the object, it's geometrical shape and reflections from another objects for divided Lidar ray profiles.
Each reflected scan point carries two information, depth and echo pulse width of the reflected scan point.

\subsection{Database}
Our Data set is 30k frames, with frame rate of 25Hz, divided into two main traces, first recorded trace is 20k frames in a Road was used for training our DNNs, second recorded trace is 10k frames in another Road was used to validate our DNNs, and in synthetic data we used different data coming from Carmaker, Gazebo, and Unity simulation environments.
\subsection{Evaluation Metrics}
One of the main advantages of our proposed sensor model is evaluating our output with a measurable metric, to see how close or far are we from realistic sensor model, however we faced a huge challenge to achieve this because in simulated environments we don't have a ground truth or how the output should look like, which lead us to use statistical evaluation metrics, and devide our evaluation KPIs to two sets, first Real Vs Real, second real Vs simulated.

The biggest problem in evaluating this problem is the association property, where how would we associate our reference is the real data, and our the coming from different simulation environments, where the nature of these two data types are different. the only way to over come this problem is to use the unsupervised evaluation techniques, where we are evaluating distribution Vs distribution, and apply evaluation on multiple stages, over all scenario evaluation(Real Vs Sim), class per class evaluation(Real Vs Sim), and box to box evaluation(Real Vs Sim).

\begin{figure}[thpb]
  \centering
  \framebox{\includegraphics[scale=0.23]{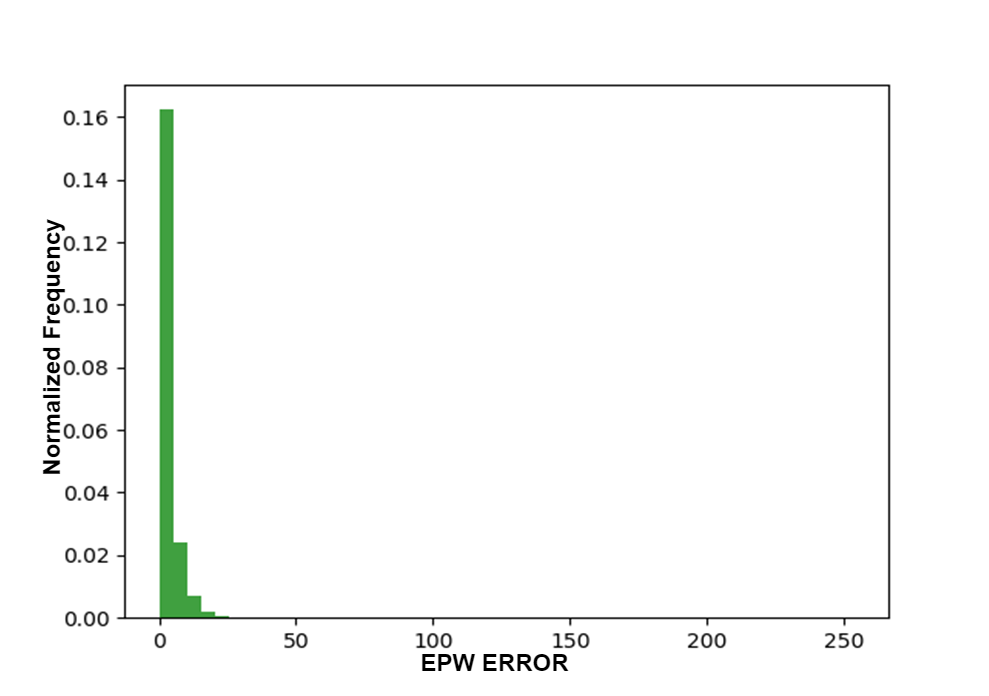}}
    \caption{\label{fig:EPWErrorHistogram} Error Histogram of inferred EPWs.}
\end{figure}

First Real Vs Real, we measure the average error of the inferred EPWs from the DNN, which was equal to \textbf{2.3058474 ns} as pulse width Error, but the number is not the best indicator to how good or bad the DNN would behave, so we plotted Error Histogram of the inferred EPWs, as shown in fig [\ref{fig:EPWErrorHistogram}].

\begin{figure}[thpb]
  \centering
    \framebox{\includegraphics[width=\columnwidth]{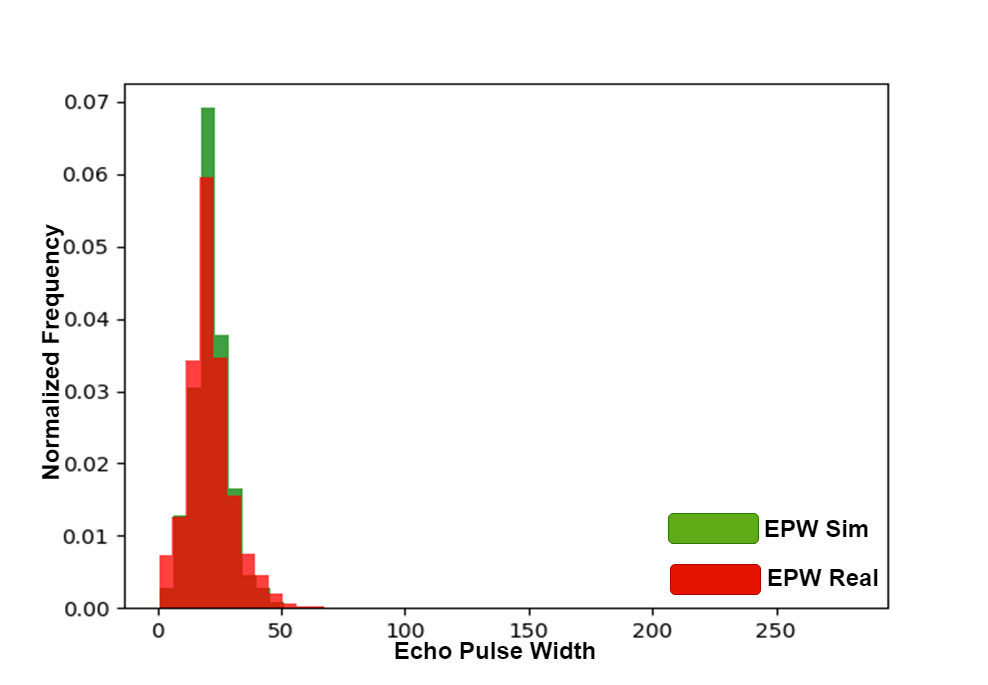}}
    \caption{\label{fig:EPWDistributionRealValidationDataHistogram} EPW evaluation from real trace Real EPW from the trace Vs Inferred EPW from the same trace.}
\end{figure}

However, if the most of the input matrix and most of the output matrix are zeros, then this wouldn’t be the best indication, so we measured this histogram of the nonzero EPWs in real Vs inferred on real data, as shown fig  [\ref{fig:EPWDistributionRealValidationDataHistogram}].

\begin{figure}[thpb]
  \centering
  \framebox{\includegraphics[width=\columnwidth]{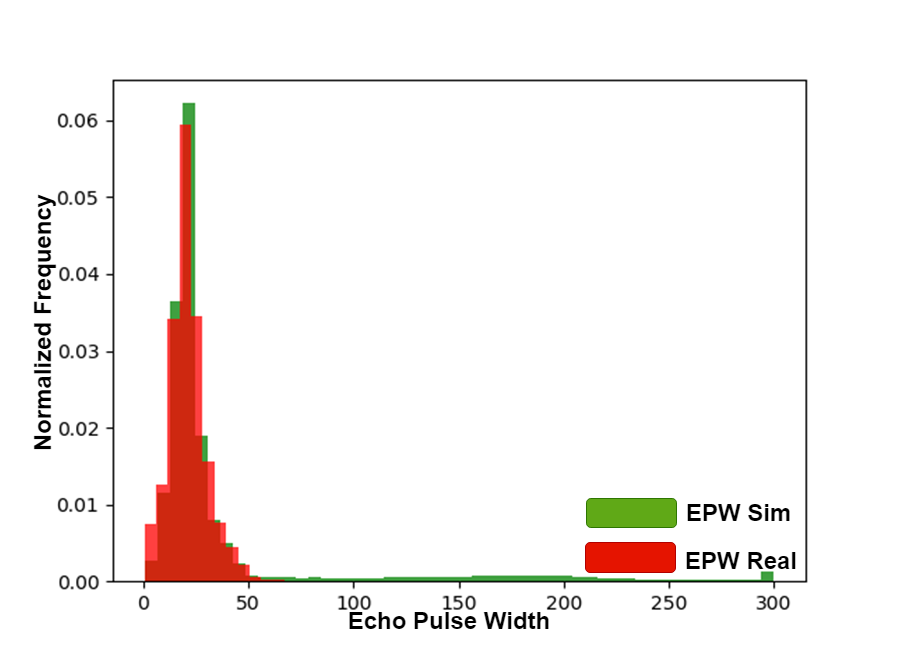}}
    \caption{\label{fig:EPWDistributionsimulationVsrealValidationDataHistogram} EPW evaluation between real trace and simulated trace that looks like the real trace, Real  EPW from Real trace Vs Inferred EPW from Simulated trace.}
\end{figure}

Second Real Vs Simulated, where we will plot the histogram of inferred EPWs from simulated environment Vs the real EPWs of real Scala gen2 Data, of a scenario that is close to the real scenario, and the results is as shown in fig [\ref{fig:EPWDistributionsimulationVsrealValidationDataHistogram}].

\begin{figure}[thpb]
  \centering
  \framebox{\includegraphics[scale=0.12]{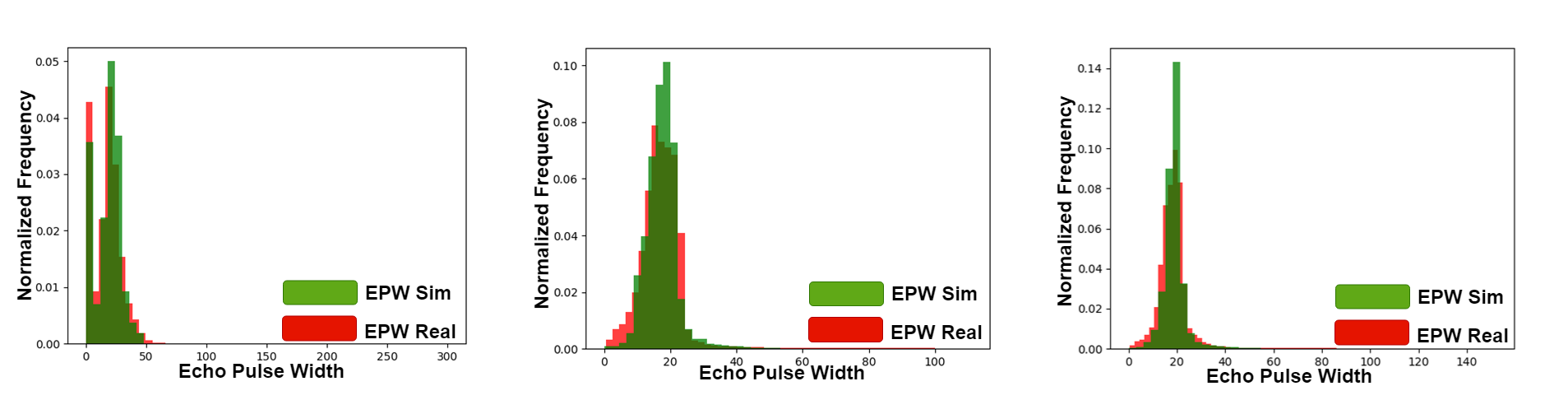}}
    \caption{\label{fig:C2CKPI} EPW Class to Class evaluation KPI, left is the None Class, middle is the Car class, right is the Truck class.}
\end{figure}

Fourth evaluation metric is the class to class evaluation metric where we evaluate each class EPW distribution in Real sensor perception Vs DNN inferred, as shown in fig [\ref{fig:C2CKPI}].
where the mean square error of the class per class evaluation metric is \textbf{1.51032 ns} error in EPW for None class, \textbf{2.31995 ns} error in EPW for Car class, and \textbf{2.89777 ns} error in EPW for Truck class.

\begin{figure}[thpb]
    \centering
      \framebox{\includegraphics[scale=0.26]{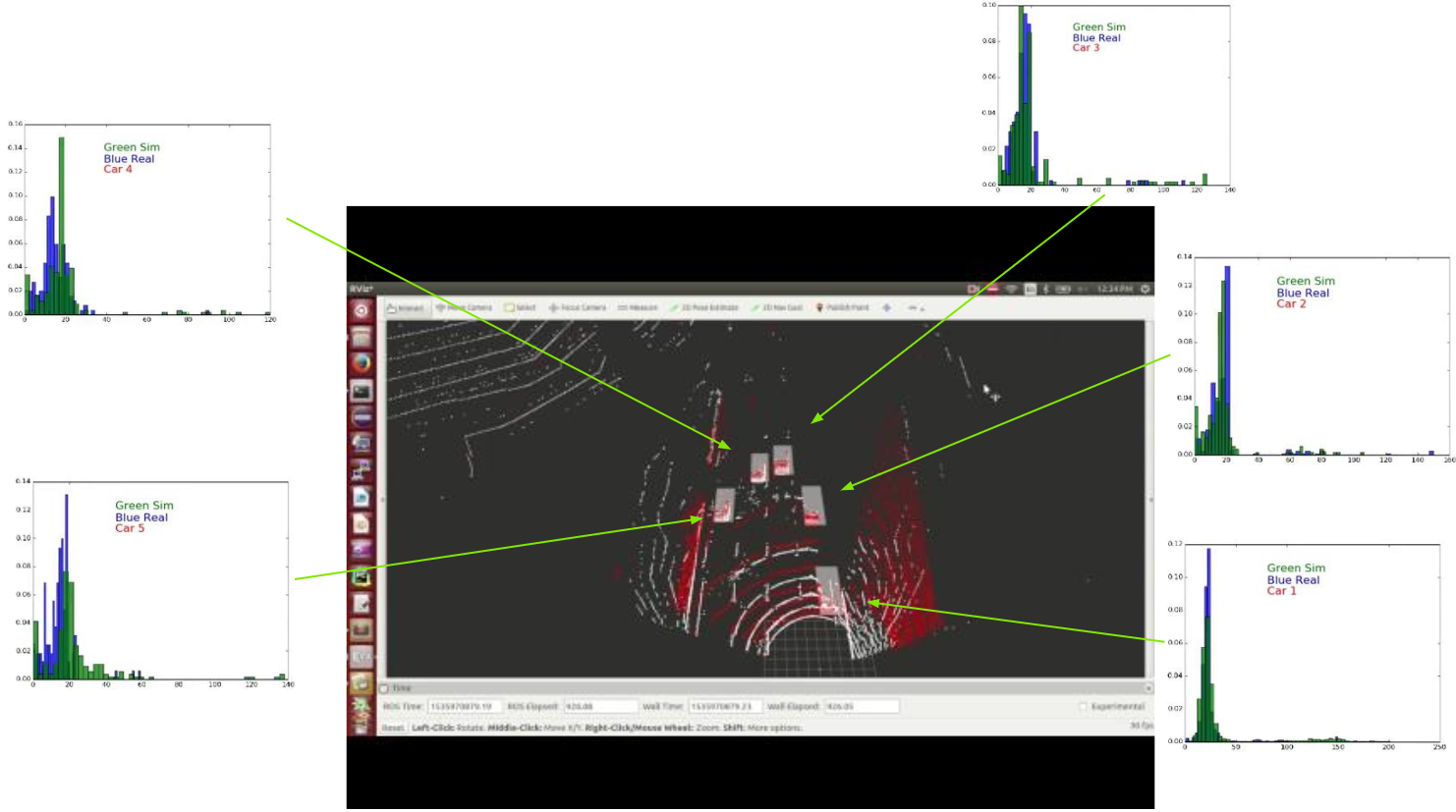}}
    \caption{\label{fig:B2BKPI} EPW vehicle(Real trace) to vehicle(Simulated trace) evaluation}
\end{figure}

Fifth evaluation metric of box to box, where we evaluate scan points within a certain oriented bounding box from real scan point perception and simulated scan point perception as shown in fig [\ref{fig:B2BKPI}].

\subsection{Results}
Different implementations were made based on different DNN architectures and their execution times and performances were bench marked, the results were as follows:

\begin{table}[h]
\label{table_results}
\begin{center}
\begin{tabular}{|c||c||c||c|}

	\hline
	EPW DNN Architecture & MSE & Accuracy & Execution time\\
	\hline
    Unet & 2.3058474 & 99.23  & 37.8ms\\
	\hline
    Unet LF & 3.0286222 & 98.99 & 13ms\\
	\hline
    tiny Unet  & 2.4942532 & 99.16 & 27ms\\
	\hline
    tiny Unet LF & 3.3249037 & 98.89 & 9.7ms\\
	\hline
    CAE & 3.5216427 & 98.82 & 17.9ms\\
	\hline
    CAE LF & 4.3697996 & 98.54 & 8.03ms\\
	\hline
\end{tabular}
\end{center}
\end{table}

where \textbf{Unet}, is the architecture proposed above, \textbf{Unet LF} is the same architecture but with half the kernel number per each convolution process, \textbf{tiny Unet} is the same Unet architecture but with only one convolution process in each Unet block instead of two convolutional processes, \textbf{tiny Unet LF} is same as tiny Unet but with half the kernel number per each convolution process, \textbf{CAE} is a simple convolution auto encoder, where the encoder is 3 down sampling convolutional processes, and decoder is 3 up sampling convolutional processes, and the \textbf{CAE LF} is the same architecture of CAE but with half it's kernel numbers per each convolutional process.

By that if accuracy is needed Unet architecture would be our choice, however if faster execution time is the target CAE LF architecture can be used, and for a tread of between good execution time and good accuracy, Tiny Unet LF would be our choice.

\subsection{Deployment}
\begin{figure}[thpb]
  \centering
        \framebox{\includegraphics[scale=0.45]{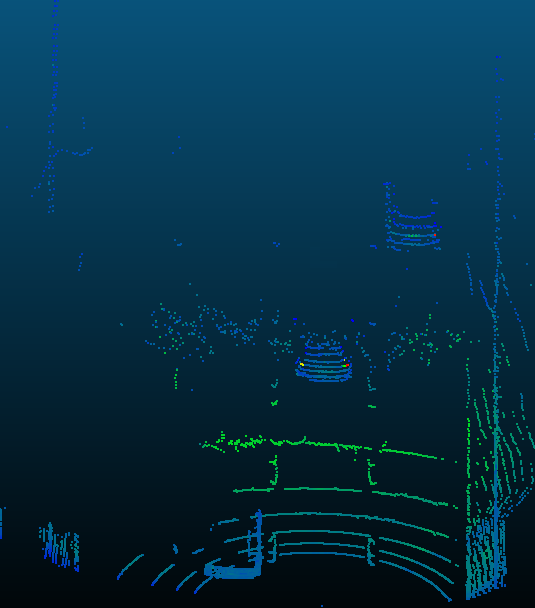}}

    \caption{\label{fig:Carmakerdemo} Carmaker deployment, where colors refers to the EPW values of each scan point}
\end{figure}

\begin{figure}[thpb]
\centering
\framebox{\includegraphics[scale=0.38]{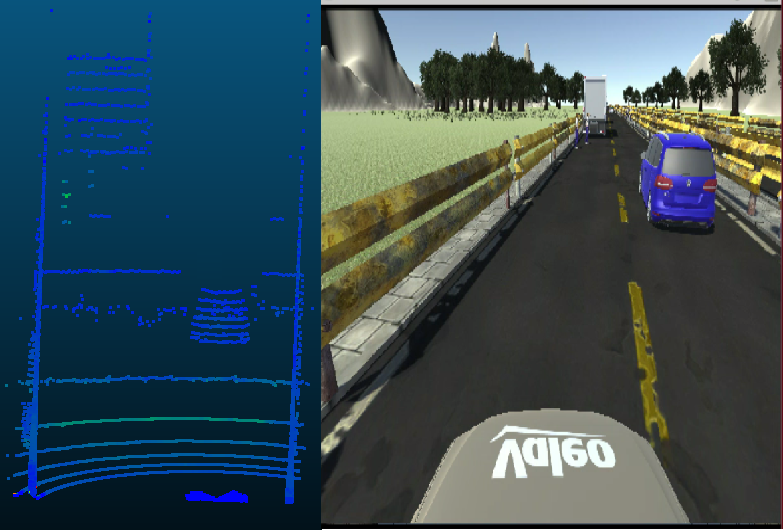}}
\caption{\label{fig:Unitydemo} Unity deployment, where colors refers to the EPW values of each scan point}
\end{figure}

\begin{figure}[thpb]
  \centering
   \framebox{\includegraphics[scale=0.24]{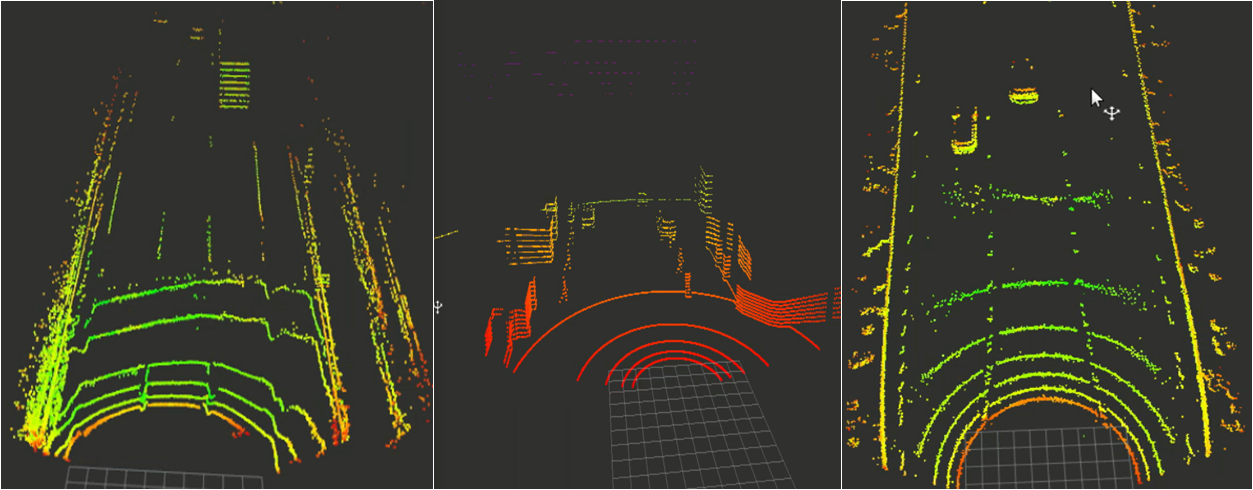}}
    \caption{\label{fig:Summary} Summary, we learn from real traces(left image), to transfer syntactic data(middle image) to be more realistic(right image).}
\end{figure}

one of the best added values of the approach that we are proposing is that it is acting as a plugin to any simulation environment, where it takes annotated point cloud as an input, and gives back Scan points perception with it's noise model and it's different physical properties. Demos for Carmaker, Unity and Gazebo are shown in fig [\ref{fig:Carmakerdemo}], fig [\ref{fig:Unitydemo}] and fig [\ref{fig:Summary}] respectively.

\section{Conclusion}
\label{sec:conc}
In this paper we have presented an approach to mimic physics of a Lidar sensor, and to represent it's sensor model in different simulation environments, such as Gazebo, Carmaker, and Unity, which have a huge impact in the developing and validation cycles, and gives a huge cost reduction, where instead of having a vehicle, and mount sensors in it, gather data, and annotate the gathered data, a cycle that costs a lot of money, and is executed in a range of one to three months, you can have the complementary solution of a simulation environment that generates for you a simulated data in a measurable way to the real data.

\section{Future Work}
\label{sec:futurework}
\begin{figure}[thpb]
\centering
\framebox{\includegraphics[scale=0.235]{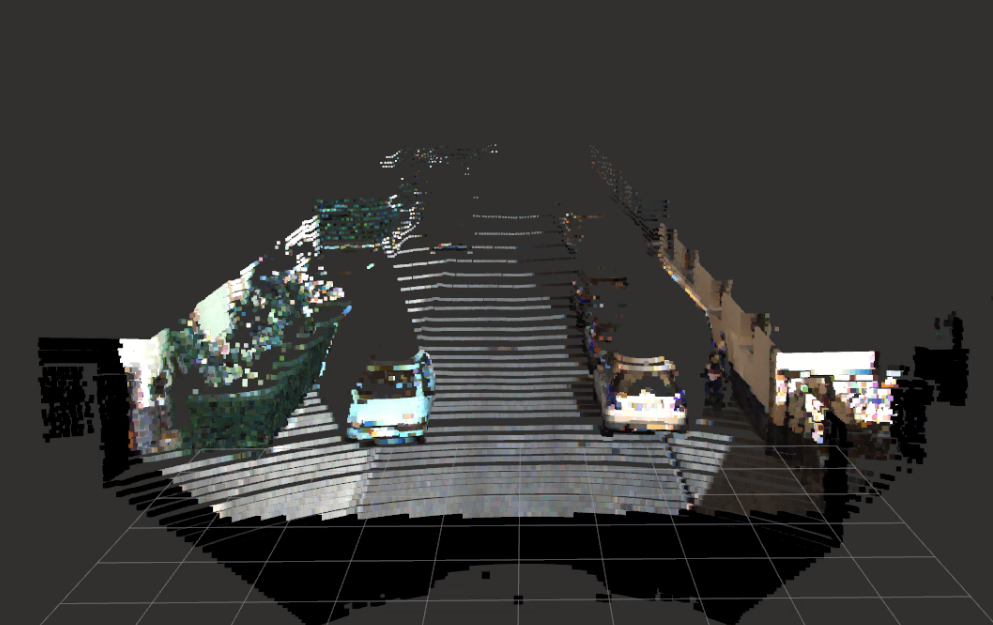}}
\caption{Camera LiDAR fused Colored point cloud.}
\label{fig:coloredPointcloud}
\end{figure}

As a future work we plan to work on using other methods in the histogram classifier, like time resolved signal modeling, and fusing these outputs instead of one out of all selection block. We investigate more quantitative KPI's to evaluate our model accuracy. Also, LiDAR-camera fusion can provide color information to scan points as shown in Fig.
\ref{fig:coloredPointcloud}. This can provide our EPW DNN and the Histogram classifier with the relation between LiDAR physical properties and object colors of each class and material. We also consider working on simulating other ADAS sensors like Radars.

{\small
\bibliographystyle{IEEEtranS}
\bibliography{eccv2018submission}
}

\end{document}